\documentclass[12pt,reqno]{amsart}
\usepackage{graphicx}
\usepackage{amscd}
\usepackage{amsmath}
\usepackage{epsfig}
\usepackage{amsfonts}
\usepackage{amssymb}

\providecommand{\U}[1]{\protect\rule{.1in}{.1in}}
\providecommand{\U}[1]{\protect\rule{.1in}{.1in}}
\allowdisplaybreaks

\theoremstyle{plain}

\numberwithin{equation}{section}

\input{tcilatex}

\begin{document}
\title[Expectation Values in Relativistic Coulomb Problems]{Expectation
Values in Relativistic Coulomb Problems}
\author{Sergei K. Suslov}
\address{School of Mathematical and Statistical Sciences and Mathematical,
Computational, and Modeling Sciences Center, Arizona State University,
Tempe, AZ 85287--1804, U.S.A.}
\email{sks@asu.edu}
\urladdr{http://hahn.la.asu.edu/\symbol{126}suslov/index.html}
\date{\today }

\begin{abstract}
We evaluate the matrix elements $\langle Or^{p}\rangle ,$ where $O$ $%
=\left\{ 1,\beta ,i\mathbf{\alpha n}\beta \right\} $ are the standard Dirac
matrix operators and the angular brackets denote the quantum-mechanical
average for the relativistic Coulomb problem, in terms of generalized
hypergeometric functions $\ _{3}F_{2}\left( 1\right) $ for all suitable
powers. Their connections with the Chebyshev and Hahn polynomials of a
discrete variable are emphasized. As a result, we derive two sets of
Pasternack-type matrix identities for these integrals, when $p\rightarrow
-p-1$ and $p\rightarrow -p-3,$ respectively. Some applications to the theory
of hydrogenlike relativistic systems are reviewed.
\end{abstract}

\maketitle

\section{Introduction}

Recent experimental progress has renewed interest in quantum electrodynamics
of atomic hydrogenlike systems. Experimentalists and theorists in atomic and
particle physics are discovering problems of common interest with new ideas
and methods. A current account of the status of this fundamental area of
quantum physics, which is more than a century old, is given in Refs.~\cite%
{Karsh01}, \cite{Kash03}, \cite{Mohr:Plun:Soff98}, and \cite{ShabYFN08}.
Exciting research topics vary from experimental testing of Quantum
Electrodynamics (QED) to fruitful training models for the bound-state
Quantum Chromodynamics and Bose--Einstein Condensation \cite{Dal:Gio:Pit:Str}%
, \cite{Karsh01}, \cite{Kash03}, \cite{Khriplovich}, \cite{Mohr:Plun:Soff98}%
, \cite{ShabYFN08}, and \cite{Yer:Ind:Shab}.\smallskip

The highly charged ions are an ideal testing ground for the strong-field
bound-state QED. They posses a strong static Coulomb field of the nucleus
and a simple electronic structure which can be accurately computed from
first principles. It is possible nowadays to make massive highly charged
ions with a strong nuclear charge and only one electron through the periodic
table up to uranium, the most highly charged ion \cite{Gum05}, \cite{Gum07}.
These systems are truly relativistic and require the Dirac wave equation as
a starting point in a detailed investigation of their spectra \cite%
{Mohr:Plun:Soff98}, \cite{ShabGreen}. The binding energy of a single $K$%
-shell electron in the electric field of a uranium nucleus corresponds to
roughly one third of the electron rest mass. For the simple hydrogen atom
the nonrelativistic Schr\"{o}dinger approximation can be used \cite{Be:Sal}%
.\smallskip

For the last decade, the two-time Green's function method of deriving formal
expressions for the energy shift of a bound-state level of high-$Z$
few-electron systems was developed \cite{ShabGreen} and numerical
calculations of QED effects in heavy ions were performed with excellent
agreement to current experimental data \cite{Gum05}, \cite{Gum07} (see \cite%
{Shab94}, \cite{Shabetal98}, \cite{ShabHyd}, \cite{Shabetal06}, \cite%
{ShabYFN08}, and \cite{Yer:Ind:Shab} and references therein for more
details). These advances motivate, among other technical things, evaluation
of the expectation values $\langle Or^{p}\rangle $ for the standard Dirac
matrix operators $O$ $=\left\{ 1,\beta ,i\mathbf{\alpha n}\beta \right\} $
between the bound-state relativistic Coulomb wave functions. Special cases
appear in calculations of the magnetic dipole hyperfine splitting, the
electric quadrupole hyperfine splitting, the anomalous Zeeman effect, and
the relativistic recoil corrections in hydrogenlike ions (see, for example, 
\cite{Adkins}, \cite{ShabHydVir}, and \cite{ShabHyd} and references
therein). We discuss convenient closed forms of these integrals in general
and derive matrix symmetry relations among them which can be useful in the
theory of relativistic Coulomb systems.\medskip

The paper is organized as follows. In the next section we review the
relativistic Coulomb wave functions and set up the notations. The
expectation values $\langle Or^{p}\rangle $ are evaluated in section~3 in
terms of the generalized hypergeometric functions $_{3}F_{2}\left( 1\right) $
for all admissible powers of $r.$ Their Pasternack-type matrix symmetry
relations are established in section~4 and recurrence relations are given in
section~5. We discuss special matrix elements and review some of their
applications in the last section. An attempt to collect the available
literature is made. The appendix~A contains definition of the generalized
hypergeometric series and proof of a required transformation identity. The
appendix~B deals with the Dirac matrices and inner product.

\section{Wave Functions for the Relativistic Coulomb Problem}

The exact solutions of the stationary Dirac equation%
\begin{equation}
H\psi =\left( c\mathbf{\alpha p}+mc^{2}\beta -Ze^{2}/r\right) \psi =E\psi
\label{sde}
\end{equation}%
for the Coulomb potential can be obtained in the spherical coordinates. The
energy levels were discovered in 1916 by Sommerfeld \cite{Somm16} from the
\textquotedblleft old\textquotedblright\ quantum theory and the
corresponding (bispinor) Dirac wave functions were found later by Darwin 
\cite{Dar} and Gordon \cite{Gor} at the early age of discovery of the
\textquotedblleft new\textquotedblright\ wave mechanics (see also \cite{Bie}
for a modern discussion of \textquotedblleft Sommerfeld's
puzzle\textquotedblright ). These classical results are nowadays included in
all textbooks on relativistic quantum mechanics, quantum field theory and
advanced texts on mathematical physics (see, for example, \cite{Akh:Ber}, 
\cite{Ber:Lif:Pit}, \cite{Be:Sal}, \cite{It:Zu}, \cite{Mes}, and \cite{Ni:Uv}%
). The end result is%
\begin{equation}
\psi =\left( 
\begin{array}{c}
\mathbf{\varphi \medskip } \\ 
\mathbf{\chi }%
\end{array}%
\right) =\left( 
\begin{array}{c}
\mathcal{Y}_{jm}^{\pm }\left( \mathbf{n}\right) \ F\left( r\right) \medskip
\\ 
i\mathcal{Y}_{jm}^{\mp }\left( \mathbf{n}\right) \ G\left( r\right)%
\end{array}%
\right) ,  \label{rc1}
\end{equation}%
where the spinor spherical harmonics $\mathcal{Y}_{jm}^{\pm }\left( \mathbf{n%
}\right) =\mathcal{Y}_{jm}^{\left( j\pm 1/2\right) }\left( \mathbf{n}\right) 
$ are given explicitly in terms of the ordinary spherical functions $%
Y_{lm}\left( \mathbf{n}\right) ,$ $\mathbf{n}=\mathbf{n}\left( \theta
,\varphi \right) =\mathbf{r}/r$ and the special Clebsch--Gordan coefficients
with the spin $1/2$ as\ follows \cite{Akh:Ber}, \cite{Ber:Lif:Pit}, \cite%
{Rose}, \cite{Var:Mos:Kher}:%
\begin{equation}
\mathcal{Y}_{jm}^{\pm }\left( \mathbf{n}\right) =\left( 
\begin{array}{c}
\mp \sqrt{\dfrac{\left( j+1/2\right) \mp \left( m-1/2\right) }{2j+\left(
1\pm 1\right) }}\ Y_{j\pm 1/2,\ m-1/2}\left( \mathbf{n}\right) \medskip \\ 
\sqrt{\dfrac{\left( j+1/2\right) \pm \left( m+1/2\right) }{2j+\left( 1\pm
1\right) }}\ Y_{j\pm 1/2,\ m+1/2}\left( \mathbf{n}\right)%
\end{array}%
\right)  \label{rc2}
\end{equation}%
with the total angular momentum $j=1/2,3/2,5/2,\ ...$ and its projection $%
m=-j,-j+1,\ ...,j-1,j$ (see also Section~VI~A of Ref.~\cite{Sus:Trey} for
the properties of the spinor spherical harmonics).\medskip\ 

The radial functions $F\left( r\right) $ and $G\left( r\right) $ can be
presented as \cite{Ni:Uv}%
\begin{eqnarray}
\left( 
\begin{array}{c}
F\left( r\right) \medskip \mathbf{\medskip \bigskip } \\ 
G\left( r\right)%
\end{array}%
\right) &=&\frac{a^{2}\beta ^{3/2}}{\nu }\sqrt{\frac{\left( \varepsilon
\kappa -\nu \right) n!}{\mu \left( \kappa -\nu \right) \Gamma \left( n+2\nu
\right) }}\ \xi ^{\nu -1}e^{-\xi /2}  \notag \\
&&\times \left( 
\begin{array}{c}
f_{1}\qquad f_{2}\mathbf{\medskip } \\ 
g_{1}\qquad g_{2}%
\end{array}%
\right) \left( 
\begin{array}{c}
\xi L_{n-1}^{2\nu +1}\left( \xi \right) \bigskip \mathbf{\medskip } \\ 
L_{n}^{2\nu -1}\left( \xi \right)%
\end{array}%
\right) .  \label{rc3}
\end{eqnarray}%
Here, $L_{k}^{\alpha }\left( \xi \right) $ are the Laguerre polynomials
given by (\ref{Laguerre}) and we use the following notations:%
\begin{eqnarray}
&&\kappa =\pm \left( j+1/2\right) ,\qquad \nu =\sqrt{\kappa ^{2}-\mu ^{2}}%
,\qquad \mu =\alpha Z=Ze^{2}/\hbar c,  \label{notations} \\
&&a=\sqrt{1-\varepsilon ^{2}},\qquad \varepsilon =E/mc^{2},\qquad \beta
=mc/\hbar =1/\lambda ,  \notag
\end{eqnarray}%
and%
\begin{equation}
\xi =2a\beta r=2\sqrt{1-\varepsilon ^{2}}\ \frac{mc}{\hbar }\ r.  \label{rc4}
\end{equation}%
The elements of $2\times 2$--transition matrix in (\ref{rc3}) are given by%
\begin{equation}
f_{1}=\frac{a\mu }{\varepsilon \kappa -\nu },\quad f_{2}=\kappa -\nu ,\quad
g_{1}=\frac{a\left( \kappa -\nu \right) }{\varepsilon \kappa -\nu },\quad
g_{2}=\mu .  \label{rc5}
\end{equation}%
This particular form of the relativistic radial functions is due to
Nikiforov and Uvarov \cite{Ni:Uv}; it is very convenient for taking the
nonrelativistic limit $c\rightarrow \infty $ (see also \cite{Sus:Trey}%
).\smallskip \smallskip

The relativistic discrete energy levels $\varepsilon =\varepsilon
_{n}=E_{n}/E_{0}$ with the rest mass energy $E_{0}=mc^{2}$ are given by the
Sommerfeld--Dirac fine structure formula%
\begin{equation}
E_{n}=\frac{mc^{2}}{\sqrt{1+\mu ^{2}/\left( n+\nu \right) ^{2}}}\ .
\label{rc6}
\end{equation}%
Here, $n=n_{r}=0,1,2,\ ...$ is the radial quantum number and $\kappa =\pm
\left( j+1/2\right) =\pm 1,\pm 2,\pm 3,\ ...\ .$ The following identities%
\begin{eqnarray}
&&\varepsilon \mu =a\left( \nu +n\right) ,\quad \varepsilon \mu +a\nu
=a\left( n+2\nu \right) ,\quad \varepsilon \mu -a\nu =an,  \label{idents} \\
&&\varepsilon ^{2}\kappa ^{2}-\nu ^{2}=a^{2}n\left( n+2\nu \right) =\mu
^{2}-a^{2}\kappa ^{2}  \notag
\end{eqnarray}%
are useful in calculation of the matrix elements below.\medskip

The familiar recurrence relations for the Laguerre polynomials allow to
present the radial functions (\ref{rc3}) in a traditional form \cite{Akh:Ber}%
, \cite{Ber:Lif:Pit}, \cite{Davis}, \cite{Mart-Rom00}, \cite{Sus:Trey} as
follows 
\begin{eqnarray}
\left( 
\begin{array}{c}
F\left( r\right) \medskip \mathbf{\medskip \bigskip } \\ 
G\left( r\right)%
\end{array}%
\right) &=&a^{2}\beta ^{3/2}\sqrt{\frac{n!}{\mu \left( \kappa -\nu \right)
\left( \varepsilon \kappa -\nu \right) \Gamma \left( n+2\nu \right) }}\ \xi
^{\nu -1}e^{-\xi /2}  \label{rc7} \\
&&\times \left( 
\begin{array}{c}
\alpha _{1}\qquad \alpha _{2}\mathbf{\medskip \medskip } \\ 
\beta _{1}\qquad \beta _{2}%
\end{array}%
\right) \left( 
\begin{array}{c}
L_{n-1}^{2\nu }\left( \xi \right) \bigskip \mathbf{\medskip } \\ 
L_{n}^{2\nu }\left( \xi \right)%
\end{array}%
\right) ,  \notag
\end{eqnarray}%
where%
\begin{equation}
\alpha _{1}=\sqrt{1+\varepsilon }\left( \left( \kappa -\nu \right) \sqrt{%
1+\varepsilon }+\mu \sqrt{1-\varepsilon }\right) ,\quad \alpha _{2}=-\sqrt{%
1+\varepsilon }\left( \left( \kappa -\nu \right) \sqrt{1+\varepsilon }-\mu 
\sqrt{1-\varepsilon }\right) ,  \label{maa}
\end{equation}%
\begin{equation}
\beta _{1}=\sqrt{1-\varepsilon }\left( \left( \kappa -\nu \right) \sqrt{%
1+\varepsilon }+\mu \sqrt{1-\varepsilon }\right) ,\quad \beta _{2}=\sqrt{%
1-\varepsilon }\left( \left( \kappa -\nu \right) \sqrt{1+\varepsilon }-\mu 
\sqrt{1-\varepsilon }\right)  \label{mbb}
\end{equation}%
and a convenient identity holds%
\begin{equation}
\left( \left( \kappa -\nu \right) \sqrt{1+\varepsilon }\pm \mu \sqrt{%
1-\varepsilon }\right) ^{2}=2\left( \kappa -\nu \right) \left( \kappa -\nu
\varepsilon \pm a\mu \right) .  \label{ident}
\end{equation}

We give the explicit form \ of the radial wave functions (\ref{rc3}) for the 
$1s_{1/2}~$state, when $n=n_{r}=0,$ $l=0,$ $j=1/2,$ and $\kappa =-1:$%
\begin{equation}
\left( 
\begin{array}{c}
F\left( r\right) \medskip \mathbf{\medskip \bigskip } \\ 
G\left( r\right)%
\end{array}%
\right) =\left( \frac{2Z}{a_{0}}\right) ^{3/2}\sqrt{\frac{\nu _{1}+1}{%
2\Gamma \left( 2\nu _{1}+1\right) }}\ \left( 
\begin{array}{c}
-1\medskip \mathbf{\medskip \bigskip } \\ 
\sqrt{\dfrac{1-\nu _{1}}{1+\nu _{1}}}%
\end{array}%
\right) \xi _{1}^{\nu _{1}-1}e^{-\xi _{1}/2}.  \label{rc6b}
\end{equation}%
Here, $\nu _{1}=\sqrt{1-\mu ^{2}}=\varepsilon _{1},$ $\xi _{1}=2\sqrt{%
1-\varepsilon _{1}^{2}}\beta r=2Z\left( r/a_{0}\right) ,$ and $a_{0}=\hbar
^{2}/me^{2}$ is the Bohr radius. One can see also \cite{Akh:Ber}, \cite%
{Ber:Lif:Pit}, \cite{Be:Sal}, \cite{Dar}, \cite{Ep76}, \cite{Gor}, \cite%
{It:Zu}, \cite{Mart-Rom00}, \cite{Mes}, and \cite{Schiff} and references
therein for more information on the relativistic Coulomb problem.

\section{Evaluation of the Matrix Elements}

We evaluate the following integrals of the radial functions:%
\begin{eqnarray}
A_{p} &=&\int_{0}^{\infty }r^{p+2}\left( F^{2}\left( r\right) +G^{2}\left(
r\right) \right) \ dr,  \label{meA} \\
B_{p} &=&\int_{0}^{\infty }r^{p+2}\left( F^{2}\left( r\right) -G^{2}\left(
r\right) \right) \ dr,  \label{meB} \\
C_{p} &=&\int_{0}^{\infty }r^{p+2}F\left( r\right) G\left( r\right) \ dr
\label{meC}
\end{eqnarray}%
in terms of generalized hypergeometric series. (Their relations with the
expectation values of the operators $\langle Or^{p}\rangle ,$ where $O$ $%
=\left\{ 1,\beta ,i\mathbf{\alpha n}\beta \right\} ,$ are discussed in the
appendix~B.) The final results with the notations from the previous section
can be presented in two different closed forms. Use of the traditional
radial functions (\ref{rc7}) results in:%
\begin{eqnarray}
&&2\mu \left( 2a\beta \right) ^{p}\ \frac{\Gamma \left( 2\nu +1\right) }{%
\Gamma \left( 2\nu +p+1\right) }\ A_{p}=2p\varepsilon an~_{3}F_{2}\left( 
\begin{array}{c}
1-n,\ -p,\ p\medskip +1 \\ 
2\nu +1,\quad 2%
\end{array}%
\right)  \label{BestA} \\
&&\qquad +\left( \mu +a\kappa \right) ~_{3}F_{2}\left( 
\begin{array}{c}
1-n,\ -p,\ p+1\medskip \\ 
2\nu +1,\quad 1%
\end{array}%
\right) +~\left( \mu -a\kappa \right) ~_{3}F_{2}\left( 
\begin{array}{c}
-n,\ -p,\ p+1\medskip \\ 
2\nu +1,\quad 1%
\end{array}%
\right) ,  \notag
\end{eqnarray}%
\begin{eqnarray}
&&2\mu \left( 2a\beta \right) ^{p}\ \frac{\Gamma \left( 2\nu +1\right) }{%
\Gamma \left( 2\nu +p+1\right) }\ B_{p}=2pan~_{3}F_{2}\left( 
\begin{array}{c}
1-n,\ -p,\ p\medskip +1 \\ 
2\nu +1,\quad 2%
\end{array}%
\right)  \label{BestB} \\
&&\qquad +\varepsilon \left( \mu +a\kappa \right) ~_{3}F_{2}\left( 
\begin{array}{c}
1-n,\ -p,\ p+1\medskip \\ 
2\nu +1,\quad 1%
\end{array}%
\right) +~\varepsilon \left( \mu -a\kappa \right) ~_{3}F_{2}\left( 
\begin{array}{c}
-n,\ -p,\ p+1\medskip \\ 
2\nu +1,\quad 1%
\end{array}%
\right) ,  \notag
\end{eqnarray}%
\begin{eqnarray}
&&4\mu \left( 2a\beta \right) ^{p}\ \frac{\Gamma \left( 2\nu +1\right) }{%
\Gamma \left( 2\nu +p+1\right) }\ C_{p}  \label{BestC} \\
&&\qquad =a\left( \mu +a\kappa \right) ~_{3}F_{2}\left( 
\begin{array}{c}
1-n,\ -p,\ p+1\medskip \\ 
2\nu +1,\quad 1%
\end{array}%
\right) -a\left( \mu -a\kappa \right) ~_{3}F_{2}\left( 
\begin{array}{c}
-n,\ -p,\ p+1\medskip \\ 
2\nu +1,\quad 1%
\end{array}%
\right) .  \notag
\end{eqnarray}%
Nikiforov and Uvarov's form (\ref{rc3}) gives the following result:%
\begin{eqnarray}
&&4\mu \nu ^{2}\left( 2a\beta \right) ^{p}\ A_{p}  \label{meAf} \\
&&\quad =a\kappa \left( \varepsilon \kappa +\nu \right) \frac{\Gamma \left(
2\nu +p+3\right) }{\Gamma \left( 2\nu +2\right) }~_{3}F_{2}\left( 
\begin{array}{c}
1-n,\ p+2,\ -p-1\medskip \\ 
2\nu +2,\quad 1%
\end{array}%
\right)  \notag \\
&&\quad -2\left( p+2\right) a^{2}\mu n\frac{\Gamma \left( 2\nu +p+2\right) }{%
\Gamma \left( 2\nu +1\right) }~_{3}F_{2}\left( 
\begin{array}{c}
1-n,\ p+2,\ -p\medskip -1 \\ 
2\nu +1,\quad 2%
\end{array}%
\right)  \notag \\
&&\quad +a\kappa \left( \varepsilon \kappa -\nu \right) \frac{\Gamma \left(
2\nu +p+1\right) }{\Gamma \left( 2\nu \right) }~_{3}F_{2}\left( 
\begin{array}{c}
-n,\ p+2,\ -p-1\medskip \\ 
2\nu ,\quad 1%
\end{array}%
\right) ,  \notag
\end{eqnarray}%
\begin{eqnarray}
&&4\mu \nu \left( 2a\beta \right) ^{p}\ B_{p}  \label{meBf} \\
&&\quad =a\left( \varepsilon \kappa +\nu \right) \frac{\Gamma \left( 2\nu
+p+3\right) }{\Gamma \left( 2\nu +2\right) }~_{3}F_{2}\left( 
\begin{array}{c}
1-n,\ p+2,\ -p-1\medskip \\ 
2\nu +2,\quad 1%
\end{array}%
\right)  \notag \\
&&\quad -a\left( \varepsilon \kappa -\nu \right) \frac{\Gamma \left( 2\nu
+p+1\right) }{\Gamma \left( 2\nu \right) }~_{3}F_{2}\left( 
\begin{array}{c}
-n,\ p+2,\ -p-1\medskip \\ 
2\nu ,\quad 1%
\end{array}%
\right) ,  \notag
\end{eqnarray}%
\begin{eqnarray}
&&8\mu \nu ^{2}\left( 2a\beta \right) ^{p}\ C_{p}  \label{meCf} \\
&&\quad =a\mu \left( \varepsilon \kappa +\nu \right) \frac{\Gamma \left(
2\nu +p+3\right) }{\Gamma \left( 2\nu +2\right) }~_{3}F_{2}\left( 
\begin{array}{c}
1-n,\ p+2,\ -p-1\medskip \\ 
2\nu +2,\quad 1%
\end{array}%
\right)  \notag \\
&&\quad -2\left( p+2\right) a^{2}\kappa n\frac{\Gamma \left( 2\nu
+p+2\right) }{\Gamma \left( 2\nu +1\right) }~_{3}F_{2}\left( 
\begin{array}{c}
1-n,\ p+2,\ -p\medskip -1 \\ 
2\nu +1,\quad 2%
\end{array}%
\right)  \notag \\
&&\quad +a\mu \left( \varepsilon \kappa -\nu \right) \frac{\Gamma \left(
2\nu +p+1\right) }{\Gamma \left( 2\nu \right) }~_{3}F_{2}\left( 
\begin{array}{c}
-n,\ p+2,\ -p-1\medskip \\ 
2\nu ,\quad 1%
\end{array}%
\right) .  \notag
\end{eqnarray}%
Here, the terminating generalized hypergeometric series $_{3}F_{2}\left(
1\right) $ are related to the Hahn and Chebyshev polynomials of a discrete
variable \cite{Ni:Su:Uv}, \cite{Sus:Trey}. (See Eqs.~(\ref{Hahn}) and (\ref%
{a1}) below, we usually omit the argument of the hypergeometric series $%
_{3}F_{2}$ if it is equal to $1.$) Two more forms occur if one takes one of
the radial wave functions from (\ref{rc3}) and another one from (\ref{rc7}).
We leave the details to the reader.\medskip

The averages of $r^{p}$ for the relativistic hydrogen atom were evaluated by
Davis \cite{Davis} in a form which is slightly different from our equations (%
\ref{BestA}) and (\ref{meAf}); see also \cite{Andrae97} and Ref.~\cite%
{Sus:Trey} for a simple proof of the second formula including evaluation of
the corresponding integral of the product of two Laguerre polynomials:%
\begin{align}
& \int_{0}^{\infty }e^{-x}x^{\alpha +s}\ L_{n}^{\alpha }\left( x\right)
L_{m}^{\beta }\left( x\right) \ dx  \label{intLaguerre} \\
& =\left( -1\right) ^{n-m}\frac{\Gamma \left( \alpha +s+1\right) \Gamma
\left( \beta +m+1\right) \Gamma \left( s+1\right) }{m!\left( n-m\right)
!\;\Gamma \left( \beta +1\right) \Gamma \left( s-n+m+1\right) }  \notag \\
& \quad \times \ ~_{3}F_{2}\left( 
\begin{array}{c}
-m,\ s+1,\ \beta -\alpha -s\medskip \\ 
\beta +1,\quad n-m+1%
\end{array}%
\right) ,\quad n\geq m.  \notag
\end{align}%
(The limit $c\rightarrow \infty $ of the integral $A_{p}$ is discussed in 
\cite{Sus:Trey}.) Equations (\ref{BestB})--(\ref{BestC}) and (\ref{meBf})--(%
\ref{meCf}), which we have not been able to find in the available
literature, can be derived in a similar fashion. It does not appear to have
been noticed that the corresponding $_{3}F_{2}$ functions can be expressed
in terms of Hahn polynomials:%
\begin{equation}
h_{n}^{\left( \alpha ,\ \beta \right) }\left( x,N\right) =\left( -1\right)
^{n}\frac{\Gamma \left( N\right) \left( \beta +1\right) _{n}}{n!\;\Gamma
\left( N-n\right) }\ _{3}F_{2}\left( 
\begin{array}{c}
-n\medskip ,\ \alpha +\beta +n+1,\ -x \\ 
\beta +1\medskip ,\quad 1-N%
\end{array}%
\right) .  \label{Hahn}
\end{equation}%
The ease of handling of these matrix elements for the discrete levels is
greatly increased if use is made of the known properties of these
polynomials \cite{Erd}, \cite{Ni:Su:Uv}, and \cite{Ni:Uv}.

For example, the difference-differentiation formulas (4.34)--(4.35) of Ref.~%
\cite{Sus:Trey} (see also (\ref{a4}) below) take the following convenient
form%
\begin{eqnarray}
&&\frac{p\left( p+1\right) }{n+2\nu }~_{3}F_{2}\left( 
\begin{array}{c}
1-n,\ -p,\ p\medskip +1 \\ 
2\nu +1,\quad 2%
\end{array}%
\right) =\frac{p\left( p+1\right) }{2\nu +1}~_{3}F_{2}\left( 
\begin{array}{c}
1-n,\ 1-p,\ p\medskip +2 \\ 
2\nu +2,\quad 2%
\end{array}%
\right)   \notag \\
&&\qquad =~_{3}F_{2}\left( 
\begin{array}{c}
-n,\ -p,\ p+1\medskip  \\ 
2\nu +1,\quad 1%
\end{array}%
\right) -~_{3}F_{2}\left( 
\begin{array}{c}
1-n,\ -p,\ p+1\medskip  \\ 
2\nu +1,\quad 1%
\end{array}%
\right)   \label{Chebyshev}
\end{eqnarray}%
in terms of the generalized hypergeometric functions. (Another proof of
these identities is given in the appendix~A.) As a result, the linear
relation holds \cite{Shab91}, \cite{Adkins}%
\begin{equation}
2\kappa \left( A_{p}-\varepsilon B_{p}\right) -\left( p+1\right) \left(
B_{p}-\varepsilon A_{p}\right) =4\mu C_{p},  \label{Linear}
\end{equation}%
and we can rewrite (\ref{BestA})--(\ref{BestB}) in the following matrix form%
\begin{eqnarray}
&&2\left( p+1\right) a\mu \left( 2a\beta \right) ^{p}\ \frac{\Gamma \left(
2\nu +1\right) }{\Gamma \left( 2\nu +p+1\right) }\ \left( 
\begin{array}{c}
A_{p}\medskip  \\ 
B_{p}%
\end{array}%
\right)   \label{MartixChebyshev} \\
&&\qquad =\left( 
\begin{array}{cc}
\gamma _{1} & \medskip \gamma _{2} \\ 
\delta _{1} & \delta _{2}%
\end{array}%
\right) \left( 
\begin{array}{c}
~_{3}F_{2}\left( 
\begin{array}{c}
1-n,\ -p,\ p+1\medskip  \\ 
2\nu +1,\quad 1%
\end{array}%
\right) \medskip  \\ 
~_{3}F_{2}\left( 
\begin{array}{c}
-n,\ -p,\ p+1\medskip  \\ 
2\nu +1,\quad 1%
\end{array}%
\right) 
\end{array}%
\right) \qquad \left( p\neq -1\right) ,  \notag
\end{eqnarray}%
where%
\begin{equation}
\gamma _{1}=\left( \mu +a\kappa \right) \left( a\left( 2\varepsilon \kappa
+p+1\right) -2\varepsilon \mu \right) ,\quad \gamma _{2}=\left( \mu -a\kappa
\right) \left( a\left( 2\varepsilon \kappa +p+1\right) +2\varepsilon \mu
\right) ,  \label{gg}
\end{equation}%
\begin{equation}
\delta _{1}=\left( \mu +a\kappa \right) \left( a\left( 2\kappa +\varepsilon
\left( p+1\right) \right) -2\mu \right) ,\quad \delta _{2}=\left( \mu
-a\kappa \right) \left( a\left( 2\kappa +\varepsilon \left( p+1\right)
\right) +2\mu \right) .  \label{dd}
\end{equation}%
This representation of integrals $A_{p}$ and $B_{p}$ involves the Chebyshev
polynomials of a discrete variable $h_{p}^{\left( 0,\ 0\right) }\left(
x,-2\nu \right) $ at $x=n,$ $n-1$ only; see also equation (\ref{BestC}) for $%
C_{p}.$ The corresponding dual Hahn polynomials \cite{Ni:Su:Uv} may be
considered as difference analogs of the Laguerre polynomials in equation (%
\ref{rc7}) for the relativistic radial functions.

\section{ Inversion Formulas}

Due to the symmetry of the hypergeometric functions in (\ref{BestA})--(\ref%
{BestC}) under the transformation $p\rightarrow -p-1,$ one gets%
\begin{equation}
A_{-p-1}=\left( 2a\beta \right) ^{2p+1}\ \frac{\Gamma \left( 2\nu -p\right) 
}{\Gamma \left( 2\nu +p+1\right) }\frac{\left( \left( 1+\varepsilon
^{2}\right) p+\varepsilon ^{2}\right) A_{p}-\left( 2p+1\right) \varepsilon
B_{p}}{\left( 1-\varepsilon ^{2}\right) p},  \label{BestInvA}
\end{equation}%
\begin{equation}
B_{-p-1}=\left( 2a\beta \right) ^{2p+1}\ \frac{\Gamma \left( 2\nu -p\right) 
}{\Gamma \left( 2\nu +p+1\right) }\frac{\left( 2p+1\right) \varepsilon
A_{p}-\left( \left( 1+\varepsilon ^{2}\right) p+1\right) B_{p}}{\left(
1-\varepsilon ^{2}\right) p},  \label{BestInvB}
\end{equation}%
\begin{equation}
C_{-p-1}=\left( 2a\beta \right) ^{2p+1}\ \frac{\Gamma \left( 2\nu -p\right) 
}{\Gamma \left( 2\nu +p+1\right) }\ C_{p}.  \label{BestInvC}
\end{equation}%
(These relations allow us to evaluate all the convergent integrals with $%
p\leq -2.$) Indeed,%
\begin{equation}
A_{-p-1}-\varepsilon B_{-p-1}=\left( 2a\beta \right) ^{2p+1}\ \frac{\Gamma
\left( 2\nu -p\right) }{\Gamma \left( 2\nu +p+1\right) }\left(
A_{p}-\varepsilon B_{p}\right) ,  \label{InvAB}
\end{equation}%
\begin{equation}
B_{-p-1}-\varepsilon A_{-p-1}=-\frac{p+1}{p}\left( 2a\beta \right) ^{2p+1}\ 
\frac{\Gamma \left( 2\nu -p\right) }{\Gamma \left( 2\nu +p+1\right) }\left(
B_{p}-\varepsilon A_{p}\right) ,  \label{InvBA}
\end{equation}%
which gives the first two equations, if $B_{p}\neq \varepsilon A_{p}$ and $%
p\neq 0,-1.$ The last one follows from (\ref{BestC}). Special cases $p=0,-1$
of (\ref{InvAB})--(\ref{InvBA}) are simply identity (\ref{ABiden}) and
Fock's virial theorem (\ref{FockVir}), respectively. In view of our formulas
(\ref{BestA})--(\ref{BestB}), equation $B_{p}=\varepsilon A_{p}$ occurs only
when $p=0$ or $n=0.$\medskip

The symmetry of the hypergeometric functions in (\ref{meAf})--(\ref{meCf})
under another reflection $p\rightarrow -p-3$ gives%
\begin{eqnarray}
A_{-p-3} &=&\left( 2a\beta \right) ^{2p+3}\ \frac{\Gamma \left( 2\nu
-p-2\right) }{\Gamma \left( 2\nu +p+3\right) }  \label{invA} \\
&&\times \left( \frac{4\mu ^{2}\left( 2p+3\right) +\left( p+2\right) \left(
4\nu ^{2}+\left( p+1\right) \left( p+2\right) \right) }{p+2}\ A_{p}\right. 
\notag \\
&&\qquad -\left. 2\kappa \left( 2p+3\right) \ B_{p}-8\kappa \mu \frac{%
2p+3^{\ }}{p+2}\ C_{p}\right) ,  \notag
\end{eqnarray}%
\begin{eqnarray}
B_{-p-3} &=&\left( 2a\beta \right) ^{2p+3}\ \frac{\Gamma \left( 2\nu
-p-2\right) }{\Gamma \left( 2\nu +p+3\right) }  \label{invB} \\
&&\times \left( -2\kappa \left( 2p+3\right) \ A_{p}+\left( 4\nu ^{2}+\left(
p+1\right) \left( p+2\right) \right) \ B_{p}\right.  \notag \\
&&\qquad +\left. 4\mu \left( 2p+3\right) \ C_{p}\right) ,  \notag
\end{eqnarray}%
\begin{eqnarray}
C_{-p-3} &=&\left( 2a\beta \right) ^{2p+3}\ \frac{\Gamma \left( 2\nu
-p-2\right) }{\Gamma \left( 2\nu +p+3\right) }  \label{invC} \\
&&\times \left( 2\kappa \mu \frac{2p+3^{\ }}{p+2}\ A_{p}-\mu \left(
2p+3\right) \ B_{p}\right.  \notag \\
&&\qquad -\left. \frac{4\mu ^{2}\left( 2p+3\right) +\left( p+1\right) \left(
4\nu ^{2}-\left( p+2\right) ^{2}\right) }{p+2}\ C_{p}\right)  \notag
\end{eqnarray}%
as a result of elementary matrix multiplications. These relations can be
used for all the convergent integrals with $p\leq -3.$ Further details are
left to the reader.\medskip

The corresponding single two-term nonrelativistic relation was found by
Pasternack \cite{Past}, \cite{PastCH} (see also \cite{Shab91} and references
therein). We have been unable to find the relativistic matrix identities (%
\ref{BestInvA})--(\ref{BestInvC}) and (\ref{invA})--(\ref{invC}) in the
available literature (see Eq.~(18) of Ref.~\cite{Andrae97} as the closest
analog).

\section{ Recurrence Relations}

A set of useful recurrence relations between the relativistic matrix
elements was derived by Shabaev \cite{Shab91} (see also \cite{Ep:Ep}, \cite%
{Vrs:Ham}, \cite{ShabHydVir}, and \cite{Adkins}) on the basis of hypervirial
theorem:%
\begin{eqnarray}
2\kappa A_{p}-\left( p+1\right) B_{p} &=&4\mu C_{p}+4\beta \varepsilon
C_{p+1},  \label{rr1} \\
2\kappa B_{p}-\left( p+1\right) A_{p} &=&4\beta C_{p+1},  \label{rr2} \\
\mu B_{p}-\left( p+1\right) C_{p} &=&\beta \left( A_{p+1}-\varepsilon
B_{p+1}\right) .  \label{rr3}
\end{eqnarray}%
Linear relation (\ref{Linear}) and convenient recurrence formulas%
\begin{eqnarray}
A_{p+1} &=&-\left( p+1\right) \frac{4\nu ^{2}\varepsilon +2\kappa \left(
p+2\right) +\varepsilon \left( p+1\right) \left( 2\kappa \varepsilon
+p+2\right) }{4\left( 1-\varepsilon ^{2}\right) \left( p+2\right) \beta \mu }%
\ A_{p}  \label{rra} \\
&&+\frac{4\mu ^{2}\left( p+2\right) +\left( p+1\right) \left( 2\kappa
\varepsilon +p+1\right) \left( 2\kappa \varepsilon +p+2\right) }{4\left(
1-\varepsilon ^{2}\right) \left( p+2\right) \beta \mu }\ B_{p},  \notag
\end{eqnarray}%
\begin{eqnarray}
B_{p+1} &=&-\left( p+1\right) \frac{4\nu ^{2}+2\kappa \varepsilon \left(
2p+3\right) +\varepsilon ^{2}\left( p+1\right) \left( p+2\right) }{4\left(
1-\varepsilon ^{2}\right) \left( p+2\right) \beta \mu }\ A_{p}  \label{rrb}
\\
&&+\frac{4\mu ^{2}\varepsilon \left( p+2\right) +\left( p+1\right) \left(
2\kappa \varepsilon +p+1\right) \left( 2\kappa +\varepsilon \left(
p+2\right) \right) }{4\left( 1-\varepsilon ^{2}\right) \left( p+2\right)
\beta \mu }\ B_{p},  \notag
\end{eqnarray}%
\begin{equation}
C_{p+1}=\frac{1}{4\mu }\left( 2\kappa +\varepsilon \left( p+2\right) \right)
\ A_{p+1}-\frac{1}{4\mu }\left( 2\kappa \varepsilon +p+2\right) \ B_{p+1}
\label{rrc}
\end{equation}%
are obtained from these equations (see \cite{Shab91}, \cite{ShabHydVir}, and 
\cite{Adkins} for more details). Their connections with the theory of
generalized hypergeometric functions will be discussed elsewhere.

\section{Special Expectation Values and Their Applications}

The Sommerfeld--Dirac formula (\ref{rc6}) is derived for a point charge
atomic nucleus with infinite mass and no internal structure (electron moving
in static Coulomb field). In reality, the electron's mass is not negligibly
small compared with the nuclear mass and one has to consider the effect of
nuclear motion on the energy levels. Actual nuclei have a finite size and
possess some internal structure, such as an internal angular momentum or
spin, a magnetic dipole moment, and a small electric quadrupole moment
associated with the spin, which also affect the energy levels. Radiative
corrections are introduced by the quantization of the electromagnetic
radiation field. (See \cite{Be:Sal}, \cite{Breit:Brown}, \cite{Shab94}, \cite%
{Shabetal98}, \cite{ShabHyd}, \cite{ShabGreen}, \cite{Yer:Ind:Shab}, and 
\cite{ShabYFN08} and references therein for more details.) Calculations of
the real energy levels of the high-$Z$ one-electron systems with the help of
the perturbation theory require special relativistic matrix elements.\medskip

From the explicit expressions (\ref{BestA})--(\ref{meCf}) one can derive the
following special matrix elements:%
\begin{eqnarray}
A_{2} &=&\left\langle r^{2}\right\rangle =\frac{5n\left( n+2\nu \right)
+4\nu ^{2}+1-\varepsilon \kappa \left( 2\varepsilon \kappa +3\right) }{%
2\left( a\beta \right) ^{2}}  \label{sp1} \\
&=&\frac{2\kappa ^{2}\varepsilon ^{4}+3\kappa \varepsilon ^{3}+\left( 3\mu
^{2}-\nu ^{2}-1\right) \varepsilon ^{2}-3\kappa \varepsilon -\nu ^{2}+1}{%
2\beta ^{2}\left( 1-\varepsilon ^{2}\right) ^{2}},  \notag
\end{eqnarray}%
\begin{equation}
A_{1}=\left\langle r\right\rangle =\frac{3\varepsilon \mu ^{2}-\kappa \left(
1-\varepsilon ^{2}\right) \left( 1+\varepsilon \kappa \right) }{2\beta \mu
\left( 1-\varepsilon ^{2}\right) },  \label{sp2}
\end{equation}%
\begin{equation}
A_{0}=\left\langle 1\right\rangle =1,  \label{sp3}
\end{equation}%
\begin{eqnarray}
A_{-1} &=&\left\langle \dfrac{1}{r}\right\rangle =\frac{\beta }{\mu \nu }%
\left( 1-\varepsilon ^{2}\right) \left( \varepsilon \nu +\mu \sqrt{%
1-\varepsilon ^{2}}\right)  \label{sp4} \\
&=&\frac{m^{2}c^{4}-E^{2}}{m^{2}c^{4}}\left( \frac{E}{Ze^{2}}+\sqrt{\frac{%
m^{2}c^{4}-E^{2}}{\hbar ^{2}c^{2}\kappa ^{2}-Z^{2}e^{4}}}\right) ,  \notag
\end{eqnarray}%
\begin{equation}
A_{-2}=\left\langle \dfrac{1}{r^{2}}\right\rangle =\frac{2a^{3}\beta
^{2}\kappa \left( 2\varepsilon \kappa -1\right) }{\mu \nu \left( 4\nu
^{2}-1\right) },  \label{sp5}
\end{equation}%
\begin{equation}
A_{-3}=\left\langle \frac{1}{r^{3}}\right\rangle =2\left( a\beta \right) ^{3}%
\frac{3\varepsilon ^{2}\kappa ^{2}-3\varepsilon \kappa -\nu ^{2}+1}{\nu
\left( \nu ^{2}-1\right) \left( 4\nu ^{2}-1\right) }.  \label{sp6}
\end{equation}%
(Note that $A_{-3}$ exists only if $\left\vert \kappa \right\vert \geq 2$ 
\cite{Shab91}.) The average distance between the electron and the nucleus $%
\overline{r}=\left\langle r\right\rangle $ is given by $A_{1}.$ The mean
square deviation of the nucleus-electron separation is $\overline{\left( r-%
\overline{r}\right) ^{2}}=A_{2}-\left( A_{1}\right) ^{2}.$ The energy
eigenvalue $\left\langle E\right\rangle ,$ mean radius $\left\langle
r\right\rangle $ and mean square radius $\left\langle r^{2}\right\rangle $
are frequently used when making comparisons of wave functions computed by
different approximation methods. The integrals $A_{1}$ and $A_{2}$ have been
evaluated in \cite{Gar:May}, \cite{Burke:Grant67}, \cite{Qiang:Shi-HaiDong},
and \cite{Sus:Trey} (see also Ref.~\cite{Andrae97} for closed-form
expressions for $\left\{ A_{p}\right\} _{p=-6}^{5}$). Matrix element $A_{-3}$
appears in calculation of the electric quadrupole hyperfine splitting \cite%
{Pyy}, \cite{Shab94}, and \cite{ShabHydVir}. Integrals $A_{p}$ are also part
of the expression for the effective electrostatic potential for the
relativistic hydrogenlike atom \cite{Sus:Trey}. 
\begin{eqnarray}
B_{2} &=&\left\langle \beta r^{2}\right\rangle =\frac{\varepsilon }{2\left(
a\beta \right) ^{2}}\left( 5n\left( n+2\nu \right) +2\nu ^{2}+1-3\varepsilon
\kappa \right)  \label{sp7} \\
&=&\varepsilon \frac{3\kappa \varepsilon ^{3}+\left( 5\mu ^{2}+3\nu
^{2}-1\right) \varepsilon ^{2}-3\kappa \varepsilon -3\nu ^{2}+1}{2\beta
^{2}\left( 1-\varepsilon ^{2}\right) ^{2}},  \notag
\end{eqnarray}%
\begin{equation}
B_{1}=\left\langle \beta r\right\rangle =\frac{3\varepsilon ^{2}\mu
^{2}-\left( 1-\varepsilon ^{2}\right) \left( \varepsilon \kappa +\nu
^{2}\right) }{2\beta \mu \left( 1-\varepsilon ^{2}\right) },  \label{sp8}
\end{equation}%
\begin{equation}
B_{0}=\left\langle \beta \right\rangle =\varepsilon =\frac{E}{mc^{2}},
\label{sp9}
\end{equation}%
\begin{equation}
B_{-1}=\left\langle \dfrac{\beta }{r}\right\rangle =\frac{\beta a^{2}}{\mu }=%
\frac{m^{2}c^{4}-E^{2}}{Ze^{2}mc^{2}},  \label{sp10}
\end{equation}%
\begin{equation}
B_{-2}=\left\langle \dfrac{\beta }{r^{2}}\right\rangle =\frac{2a^{3}\beta
^{2}\left( 2\nu ^{2}-\varepsilon \kappa \right) }{\mu \nu \left( 4\nu
^{2}-1\right) },  \label{sp11}
\end{equation}%
\begin{equation}
B_{-3}=\left\langle \frac{\beta }{r^{3}}\right\rangle =2\left( a\beta
\right) ^{3}\varepsilon \frac{1+2\nu ^{2}-3\varepsilon \kappa }{\nu \left(
\nu ^{2}-1\right) \left( 4\nu ^{2}-1\right) }.  \label{sp12}
\end{equation}%
The integral $B_{0}$ appears in the virial theorem for the Dirac equation in
a Coulomb field,%
\begin{equation}
E=mc^{2}\left\langle \beta \right\rangle ,  \label{FockVir}
\end{equation}%
established by Fock \cite{FockVir} and then developed by many authors (see 
\cite{Breit:Brown}, \cite{Brown50}, \cite{Rose:Wel}, \cite{March}, \cite%
{Schect:Good57}, \cite{Dorling70}, \cite{McKinley}, \cite{RoseElec}, \cite%
{Ep:Ep}, \cite{Rosicky:Mark75}, \cite{Fri:Neg}, \cite{Good:Drake}, \cite%
{Shab91}, and \cite{ShabHydVir} and references therein). Relation (\ref%
{FockVir}) can also be obtained with the help of the Hellmann--Feynman
theorem,%
\begin{equation}
\frac{\partial E}{\partial \lambda }=\left\langle \frac{\partial H}{\partial
\lambda }\right\rangle  \label{HellFeyn}
\end{equation}%
(see \cite{Ep:Ep}, \cite{McKinley}, \cite{Balas84}, and \cite{Balas90} and
references therein), if applied to the mass parameter \cite{Adkins}, \cite%
{ShabHydVir}. This theorem implies two more relations%
\begin{equation}
\frac{\partial E}{\partial Z}=-e^{2}\left\langle \dfrac{1}{r}\right\rangle
=-e^{2}A_{-1},\qquad \frac{\partial E}{\partial \kappa }=2\hbar cC_{-1}.
\label{spHFT}
\end{equation}%
The following identities hold%
\begin{equation}
A_{-1}-\varepsilon B_{-1}=\frac{a^{3}\beta }{\nu }=\frac{1}{\beta }\left(
\mu B_{-2}+C_{-2}\right)  \label{ABiden}
\end{equation}%
by (\ref{rr3}). The integral $B_{-1}$ is evaluated in \cite{Breit:Brown} and 
$A_{-1},$ $A_{-2},$ $B_{-2},$ $C_{-2},$ and $A_{-3}$ are given in \cite%
{Shab91} (see also \cite{ShabHydVir}). \medskip

The relativistic recoil corrections to the energy levels, when nuclear
motion is taken into consideration, require matrix elements $A_{-2},$ $%
B_{-1} $ and $C_{-2}$ (see \cite{Breit:Brown}, \cite{Shabetal98}, \cite%
{ShabHyd}, and \cite{Adkins} and references therein).%
\begin{eqnarray}
C_{2} &=&\frac{\kappa a^{2}\left( 3n\left( n+2\nu \right) +2\nu
^{2}+1\right) -3\mu ^{2}\varepsilon }{4\mu \left( a\beta \right) ^{2}}
\label{sp13} \\
&=&\frac{\kappa \left( 1-\varepsilon ^{2}\right) \left( 1-\nu ^{2}\right)
+3\varepsilon \mu ^{2}\left( \varepsilon \kappa -1\right) }{4\mu \beta
^{2}\left( 1-\varepsilon ^{2}\right) },  \notag
\end{eqnarray}%
\begin{equation}
C_{1}=\frac{2\varepsilon \kappa -1}{4\beta }=\frac{\hbar }{4m^{2}c^{3}}%
\left( 2\kappa E-mc^{2}\right) ,  \label{sp14}
\end{equation}%
\begin{equation}
C_{0}=\frac{\kappa }{2\mu }\left( 1-\varepsilon ^{2}\right) =\frac{\hbar
c\kappa }{2Ze^{2}}\frac{m^{2}c^{4}-E^{2}}{m^{2}c^{4}},  \label{sp15}
\end{equation}%
\begin{eqnarray}
C_{-1} &=&\frac{\kappa }{2\mu \nu }a^{3}\beta =\frac{a\beta }{\nu }C_{0}
\label{sp16} \\
&=&\frac{\hbar \kappa }{2Ze^{2}m^{2}c^{3}}\frac{\left(
m^{2}c^{4}-E^{2}\right) ^{3/2}}{\left( \hbar ^{2}c^{2}\kappa
^{2}-Z^{2}e^{4}\right) ^{1/2}},  \notag
\end{eqnarray}%
\begin{equation}
C_{-2}=\frac{a^{3}\beta ^{2}\left( 2\varepsilon \kappa -1\right) }{\nu
\left( 4\nu ^{2}-1\right) }=\frac{4\left( a\beta \right) ^{3}C_{1}}{\nu
\left( 4\nu ^{2}-1\right) },  \label{sp17}
\end{equation}%
\begin{equation}
C_{-3}=\left( a\beta \right) ^{3}\frac{\kappa \left( 1-\varepsilon
^{2}\right) \left( 1-\nu ^{2}\right) +3\varepsilon \mu ^{2}\left(
\varepsilon \kappa -1\right) }{\mu \nu \left( \nu ^{2}-1\right) \left( 4\nu
^{2}-1\right) }=\frac{4\left( a\beta \right) ^{5}C_{2}}{\nu \left( \nu
^{2}-1\right) \left( 4\nu ^{2}-1\right) }.  \label{sp18}
\end{equation}%
The integrals $C_{0},$ $C_{1},$ and $B_{-1}$ are computed in \cite%
{Good:Drake}. In view of (\ref{sp10}) and (\ref{sp15}), respectively (\ref%
{sp5}) and (\ref{sp17}), the following simple relations hold%
\begin{equation}
C_{0}=\frac{\kappa }{2\beta }B_{-1},\qquad A_{-2}=\frac{2\kappa }{\mu }%
C_{-2}=\frac{8\left( a\beta \right) ^{3}\kappa }{\mu \nu \left( 4\nu
^{2}-1\right) }C_{1}.  \label{sp19}
\end{equation}%
The last but one was originally found in \cite{Brown50}.\medskip

The integral $C_{1}$ occurs in calculations of the bound-electron $g$ factor
(the anomalous Zeeman effect in the presence of an external homogeneous
static magnetic field) \cite{Margenau40}, \cite{RoseElec}, \cite{WongYeh83}, 
\cite{ShabHydVir}, and \cite{Shabetal06}. The matrix element $C_{-1}$ has
also been found by the Hellmann--Feynman theorem (\ref{spHFT}). The integral 
$C_{-2}$ appears in calculation of the magnetic dipole hyperfine splitting 
\cite{Breit30}, \cite{Brown50}, \cite{McKee69}, \cite{Fri:Neg}, \cite{Pyy}, 
\cite{RoseElec}, and \cite{Shab94}.\medskip

The author hopes that the rest of matrix elements will also be useful in the
current theory of hydrogenlike heavy ions and other exotic relativistic
Coulomb systems. Professor Shabaev kindly pointed out that the formulas
derived in this paper can be used in calculations with hydrogenlike wave
functions where a high precision is required.\medskip

In Table~1, we list the expectation values for the $1s_{1/2}$ state, when $%
n=n_{r}=0,$ $l=0,$ $j=1/2,$ and $\kappa =-1.$ The corresponding radial wave
functions are given by Eq.~(\ref{rc6b}).\medskip

Table~1. Expectation values for the $1s_{1/2}$ state.\medskip

\negthinspace \negthinspace 
\begin{tabular}{|c|c|c|c|}
\hline
$p$ & $A_{p}$ & $B_{p}$ & $C_{p}$ \\ \hline
$2$ & $\!\!\dfrac{1}{2}\left( \dfrac{a_{0}}{Z}\right) ^{2}\left( \nu
_{1}+1\right) \left( 2\nu _{1}+1\right) $ & $\!\dfrac{1}{2}\left( \dfrac{%
a_{0}}{Z}\right) ^{2}\nu _{1}\left( \nu _{1}+1\right) \left( 2\nu
_{1}+1\right) \!$ & $\!-\dfrac{\lambda a_{0}}{4Z}\left( \nu _{1}+1\right)
\left( 2\nu _{1}+1\right) $ \\ \hline
$1$ & $\dfrac{a_{0}}{2Z}\left( 2\nu _{1}+1\right) $ & $\dfrac{a_{0}}{2Z}\nu
_{1}\left( 2\nu _{1}+1\right) $ & $-\dfrac{\lambda }{4}\left( 2\nu
_{1}+1\right) $ \\ \hline
$0$ & $1$ & $\nu _{1}$ & $-\dfrac{\lambda Z}{2a_{0}}$ \\ \hline
$\!-1$ & $\dfrac{Z}{a_{0}\nu _{1}}$ & $\dfrac{Z}{a_{0}}$ & $-\left( \dfrac{Z%
}{a_{0}}\right) ^{2}\dfrac{\lambda }{2\nu _{1}}$ \\ \hline
$\!-2$ & $\left( \dfrac{Z}{a_{0}}\right) ^{2}\dfrac{2}{\nu _{1}\left( 2\nu
_{1}-1\right) }$ & $\left( \dfrac{Z}{a_{0}}\right) ^{2}\dfrac{2}{\left( 2\nu
_{1}-1\right) }$ & $-\left( \dfrac{Z}{a_{0}}\right) ^{3}\dfrac{\lambda }{\nu
_{1}\left( 2\nu _{1}-1\right) }$ \\ \hline
$\!-3$ & $\!\left( \dfrac{Z}{a_{0}}\right) ^{3}\dfrac{2}{\nu _{1}\left( \nu
_{1}-1\right) \left( 2\nu _{1}-1\right) }\!$ & $\left( \dfrac{Z}{a_{0}}%
\right) ^{3}\dfrac{2}{\left( \nu _{1}-1\right) \left( 2\nu _{1}-1\right) }\!$
& $\!-\left( \dfrac{Z}{a_{0}}\right) ^{4}\dfrac{\lambda }{\nu _{1}\left( \nu
_{1}-1\right) \left( 2\nu _{1}-1\right) }\!\!$ \\ \hline
\end{tabular}%
\medskip

\noindent In the table, $\varepsilon _{1}=\nu _{1}=\sqrt{1-\mu ^{2}}=\sqrt{%
1-\left( \alpha Z\right) ^{2}},$ $\alpha =e^{2}/\hbar c$ is the Sommerfeld
fine structure constant, $a_{0}=\hbar ^{2}/me^{2}$ is the Bohr radius, and $%
\lambda =\hbar /mc$ is the Compton wavelength. The relations%
\begin{equation}
B_{p}=\varepsilon _{1}\ A_{p},\qquad C_{p}=-\frac{\lambda Z}{2a_{0}}\
A_{p},\qquad A_{p}=\left( \frac{a_{0}}{2Z}\right) ^{p}\ \frac{\Gamma \left(
2\nu _{1}+p+1\right) }{\Gamma \left( 2\nu _{1}+1\right) }  \label{linground}
\end{equation}%
(for all the suitable integers $p>-2\nu _{1}-1>-3$) follow directly from (%
\ref{BestA}), (\ref{BestB}) and (\ref{meAf}), (\ref{meCf}). (The formal
expressions for $A_{-3},$ $B_{-3},$ and $C_{-3},$ when the integrals
diverge, are included into the table for \textquotedblleft
completeness\textquotedblright ; see Ref.~\cite{Adkins} for more details.)
The reflection relation (\ref{BestInvC}) holds for all the convergent
integrals $A_{p},$ $B_{p},$ and $C_{p}.$\medskip

\noindent \textbf{Acknowledgment.\/} The author is grateful to Carlos
Castillo-Ch\'{a}vez for support and encouragement. I thank
Vladimir~M.~Shabaev for important comments and for pointing out Ref.~\cite%
{Andrae97} to my attention. The author is grateful to Dirk Andrae for
valuable suggestions and list of typos. Referees' suggestions are also very
appreciated.

\appendix

\section{Generalized Hypergeometric Series}

The generalized hypergeometric series is defined as follows \cite{Ba}, \cite%
{Erd} 
\begin{eqnarray}
&&\ _{p}F_{q}\left( a_{1},\ a_{2},\ ...\ ,\ a_{p};\;b_{1},\ b_{2},\ ...\ ,\
b_{q};\ z\right)  \label{a1} \\
&&\quad =\ _{p}F_{q}\left( 
\begin{array}{c}
a_{1},\ a_{2},\ ...\ ,\ a_{p}\medskip \\ 
b_{1},\ b_{2},\ ...\ ,\ b_{q}\medskip%
\end{array}%
;\ z\right) =\sum_{n=0}^{\infty }\frac{\left( a_{1}\right) _{n}\left(
a_{2}\right) _{n}...\left( a_{p}\right) _{n}\ z^{n}}{\left( b_{1}\right)
_{n}\left( b_{2}\right) _{n}...\left( b_{q}\right) _{n}n!},  \notag
\end{eqnarray}%
where $\left( a\right) _{n}=a\left( a+1\right) ...\left( a+n-1\right)
=\Gamma \left( a+n\right) /\Gamma \left( a\right) .$ In this paper we always
have $p=3,$ $q=2,$ $z=1,$ and $a_{1}$ is a negative integer when the series
terminates. The Laguerre polynomials are given by \cite{Erd}, \cite{Ni:Su:Uv}%
, \cite{Ni:Uv}:%
\begin{equation}
L_{n}^{\alpha }\left( x\right) =\frac{\Gamma \left( \alpha +n+1\right) }{%
n!\;\Gamma \left( \alpha +1\right) }\ _{1}F_{1}\left( 
\begin{array}{c}
-n\medskip \\ 
\alpha +1\medskip%
\end{array}%
;\ x\right) .  \label{Laguerre}
\end{equation}%
\smallskip

The required identity (\ref{Chebyshev})\ can be derived from the theory of
classical polynomials in the following fashion. Let us start from the
difference equation for the Hahn polynomials $y_{m}=h_{m}^{\left( \alpha ,\
\beta \right) }\left( x,N\right) $ \cite{Ni:Su:Uv}:%
\begin{equation}
\left( \sigma \left( x\right) \nabla +\tau \left( x\right) \right) \Delta
y_{m}+\lambda _{m}y_{m}=0,  \label{a2}
\end{equation}%
where $\Delta f\left( x\right) =\nabla f\left( x+1\right) =f\left(
x+1\right) -f\left( x\right) $ and%
\begin{eqnarray}
\sigma \left( x\right) &=&x\left( \alpha +N-x\right) ,  \label{a3} \\
\tau \left( x\right) &=&\left( \beta +1\right) \left( N-1\right) -\left(
\alpha +\beta +2\right) x,  \notag \\
\lambda _{m} &=&m\left( \alpha +\beta +m+1\right) ,  \notag
\end{eqnarray}%
and use the familiar difference-differentiation formula:%
\begin{equation}
\Delta h_{m}^{\left( \alpha ,\;\beta \right) }\left( x,N\right) =\left(
\alpha +\beta +m+1\right) h_{m-1}^{\left( \alpha +1,\;\beta +1\right)
}\left( x,N-1\right) .  \label{a4}
\end{equation}%
As a result,%
\begin{equation}
\left( \sigma \left( x\right) \nabla +\tau \left( x\right) \right)
h_{m-1}^{\left( \alpha +1,\;\beta +1\right) }\left( x,N-1\right)
+mh_{m}^{\left( \alpha ,\;\beta \right) }\left( x,N\right) =0.  \label{a5}
\end{equation}%
Letting $\alpha =\beta $ and $\beta \rightarrow -1,$ one gets%
\begin{eqnarray}
&&x\left( N-x-1\right) \nabla h_{m-1}^{\left( 0,\;0\right) }\left(
x,N-1\right) =-m\lim_{\beta \rightarrow -1}h_{m}^{\left( \beta ,\;\beta
\right) }\left( x,N\right)  \label{a6} \\
&&\qquad =\left( -1\right) ^{m}m\left( m-1\right) \frac{\Gamma \left(
N-1\right) }{\Gamma \left( N-m\right) }x\ _{3}F_{2}\left( 
\begin{array}{c}
1-m\medskip ,\ m,\ 1-x \\ 
2\medskip ,\quad 2-N%
\end{array}%
\right)  \notag
\end{eqnarray}%
by (\ref{Hahn}). The last identity takes the form (\ref{Chebyshev}), if the
Chebyshev polynomials of a discrete variable $h_{m-1}^{\left( 0,\;0\right)
}\left( x,N-1\right) $ are replaced by the corresponding generalized
hypergeometric functions. (Use of (\ref{a4}) in (\ref{a6}) gives the special 
$_{3}F_{2}$ transformation.)

\section{Dirac Matrices and Inner Product}

We use the standard representations of the Dirac and Pauli matrices:%
\begin{equation}
\mathbf{\alpha }=\left( 
\begin{array}{cc}
\mathbf{0} & \mathbf{\sigma } \\ 
\mathbf{\sigma } & \mathbf{0}%
\end{array}%
\right) ,\qquad \beta =\left( 
\begin{array}{cc}
\mathbf{1} & \mathbf{0} \\ 
\mathbf{0} & -\mathbf{1}%
\end{array}%
\right) ,  \label{b1}
\end{equation}%
\begin{equation}
\sigma _{1}=\left( 
\begin{array}{cc}
0 & 1 \\ 
1 & 0%
\end{array}%
\right) ,\qquad \sigma _{2}=\left( 
\begin{array}{cc}
0 & -i \\ 
i & 0%
\end{array}%
\right) ,\qquad \sigma _{3}=\left( 
\begin{array}{cc}
1 & 0 \\ 
0 & -1%
\end{array}%
\right)   \label{b2}
\end{equation}%
with%
\begin{equation}
\mathbf{0}=\left( 
\begin{array}{cc}
0 & 0 \\ 
0 & 0%
\end{array}%
\right) ,\qquad \mathbf{1}=\left( 
\begin{array}{cc}
1 & 0 \\ 
0 & 1%
\end{array}%
\right) .  \label{b3}
\end{equation}%
The inner product of two Dirac (bispinor) wave functions%
\begin{equation}
\psi =\left( 
\begin{array}{c}
\mathbf{u}_{1} \\ 
\mathbf{v}_{1}%
\end{array}%
\right) =\left( 
\begin{array}{c}
\psi _{1} \\ 
\psi _{2} \\ 
\psi _{3} \\ 
\psi _{4}%
\end{array}%
\right) ,\qquad \phi =\left( 
\begin{array}{c}
\mathbf{u}_{2} \\ 
\mathbf{v}_{2}%
\end{array}%
\right) =\left( 
\begin{array}{c}
\phi _{1} \\ 
\phi _{2} \\ 
\phi _{3} \\ 
\phi _{4}%
\end{array}%
\right)   \label{b4}
\end{equation}%
is defined as a scalar quantity%
\begin{eqnarray}
\left\langle \psi ,\ \phi \right\rangle  &=&\int_{\mathbf{R}^{3}}\psi
^{\dagger }\phi \ dv=\int_{\mathbf{R}^{3}}\left( \mathbf{u}_{1}^{\dagger }{}%
\mathbf{u}_{2}+\mathbf{v}_{1}^{\dagger }{}\mathbf{v}_{2}\right) \ dv
\label{b5} \\
&=&\int_{\mathbf{R}^{3}}\left( \psi _{1}^{\ast }\phi _{1}+\psi _{2}^{\ast
}\phi _{2}+\psi _{3}^{\ast }\phi _{3}+\psi _{4}^{\ast }\phi _{4}\right) \ dv
\notag
\end{eqnarray}%
and the corresponding expectation values of a matrix operator $A$ are given
by%
\begin{equation}
\langle A\rangle =\left\langle \psi ,\ A\psi \right\rangle .  \label{b6}
\end{equation}%
From this definition one gets%
\begin{equation}
\langle r^{p}\rangle =A_{p},\qquad \langle \beta r^{p}\rangle =B_{p},\qquad
\langle i\mathbf{\alpha n}\beta r^{p}\rangle =-2C_{p},  \label{b7}
\end{equation}%
where the integrals $A_{p},$ $B_{p},$ and $C_{p}$ are given by (\ref{meA})--(%
\ref{meC}), respectively.\medskip 

Indeed, the first relation is derived, for example, in Ref.~\cite{Sus:Trey}
and the second one can be obtained by integrating the identity%
\begin{eqnarray}
r^{p}\psi ^{\dagger }\beta \psi  &=&r^{p}\left( \mathbf{\varphi }^{\dagger
},\ \mathbf{\chi }^{\dagger }\right) \left( 
\begin{array}{cc}
\mathbf{1} & \mathbf{0} \\ 
\mathbf{0} & -\mathbf{1}%
\end{array}%
\right) \ \left( 
\begin{array}{c}
\mathbf{\varphi \medskip } \\ 
\mathbf{\chi }%
\end{array}%
\right) =r^{p}\left( \mathbf{\varphi }^{\dagger },\ \mathbf{\chi }^{\dagger
}\right) \ \left( 
\begin{array}{c}
\mathbf{\varphi \medskip } \\ 
-\mathbf{\chi }%
\end{array}%
\right)   \label{b8} \\
&=&r^{p}\left( \mathbf{\varphi }^{\dagger }\mathbf{\varphi -\chi }^{\dagger }%
\mathbf{\chi }\right) =r^{p}\left( \mathcal{Y^{\dagger }Y}\right) \left(
F^{2}-G^{2}\right)   \notag
\end{eqnarray}
(we leave details to the reader) in a similar fashion.\medskip 

In the last case, we start from the matrix identity%
\begin{equation}
\left( \mathbf{\alpha n}\right) \beta \psi =\left( 
\begin{array}{cc}
\mathbf{0} & \mathbf{\sigma n} \\ 
\mathbf{\sigma n} & \mathbf{0}%
\end{array}%
\right) \left( 
\begin{array}{c}
\mathbf{\varphi \medskip } \\ 
-\mathbf{\chi }%
\end{array}%
\right) =\left( 
\begin{array}{c}
-\left( \mathbf{\sigma n}\right) \mathbf{\ \chi \medskip } \\ 
\left( \mathbf{\sigma n}\right) \mathbf{\ \varphi }%
\end{array}%
\right)   \label{b9}
\end{equation}%
and use the Ansatz \cite{Sus:Trey} 
\begin{equation}
\mathbf{\varphi }=\mathbf{\varphi }\left( \mathbf{r}\right) =\mathcal{Y}%
\left( \mathbf{n}\right) \ F\left( r\right) ,\qquad \mathbf{\chi }=\mathbf{%
\chi }\left( \mathbf{r}\right) =-i\left( \left( \mathbf{\sigma n}\right) 
\mathcal{Y}\left( \mathbf{n}\right) \right) \ G\left( r\right) ,
\label{Ansatz}
\end{equation}%
where $\mathbf{n}=\mathbf{r}/r$ and $\mathcal{Y}=\mathcal{Y}_{jm}^{\pm
}\left( \mathbf{n}\right) $ are the spinor spherical harmonics given by (\ref%
{rc2}). As a result,%
\begin{eqnarray}
&&ir^{p}\psi \mathcal{^{\dagger }}\left( \left( \mathbf{\alpha n}\right)
\beta \psi \right)   \label{b10} \\
&&\qquad =ir^{p}\left( \mathbf{\varphi }^{\dagger },\ \mathbf{\chi }%
^{\dagger }\right) \left( 
\begin{array}{c}
-\left( \mathbf{\sigma n}\right) \mathbf{\ \chi \medskip } \\ 
\left( \mathbf{\sigma n}\right) \mathbf{\ \varphi }%
\end{array}%
\right) =ir^{p}\left( F\mathcal{Y^{\dagger }},\ iG\mathcal{Y^{\dagger }}%
\left( \mathbf{\sigma n}\right) \right) \left( 
\begin{array}{c}
i\mathcal{Y}G\mathbf{\medskip } \\ 
\left( \mathbf{\sigma n}\right) \mathcal{Y}F%
\end{array}%
\right)   \notag \\
&&\qquad =-r^{p}\left( \mathcal{Y^{\dagger }Y}\right) FG-r^{p}\left( 
\mathcal{Y^{\dagger }}\left( \mathbf{\sigma n}\right) ^{2}\mathcal{Y}\right)
FG=-2r^{p}\left( \mathcal{Y^{\dagger }Y}\right) FG  \notag
\end{eqnarray}%
with the help of the familiar identity $\left( \mathbf{\sigma n}\right) ^{2}=%
\mathbf{n}^{2}=\mathbf{1}.$ Integration over $\mathbf{R}^{3}$ in the
spherical coordinates completes the proof.

\end{document}